\documentclass[reprint,twocolumn,superscriptaddress,preprintnumbers,amsmath,amssymb,aps,prd,floatfix]{revtex4-2}

\usepackage[utf8]{inputenc}
\usepackage{enumerate}

\usepackage{mathtools}
\usepackage{amsfonts}
\usepackage{mathrsfs}
\usepackage{bm}
\usepackage[normalem]{ulem}
\usepackage{bbold}
\usepackage{braket}
\usepackage{slashed}
\usepackage{dsfont}
\usepackage{float}
\usepackage{dcolumn}% Align table columns on decimal point
\usepackage{amssymb,amsmath}
\usepackage{titlesec}

\usepackage{graphicx}
\usepackage{color}
\usepackage{array}

\usepackage{placeins}
\usepackage{booktabs}
\usepackage{xspace}
\usepackage{hyperref}
\usepackage[nameinlink]{cleveref}
\usepackage{bookmark}
\usepackage{units}

\def\Eq#1{Eq.~\labelcref{#1}}

\def\Fig#1{Fig.~\labelcref{#1}}

\def\sec#1{Sec.~\labelcref{#1}}

\setkeys{Gin}{width=0.48\textwidth}

\usepackage{booktabs}
\usepackage{multirow}

\newcolumntype{C}{>{$}c<{$}}
\AtBeginDocument{
	\heavyrulewidth=.08em
	\lightrulewidth=.05em
	\cmidrulewidth=.03em
	\belowrulesep=.65ex
	\belowbottomsep=0pt
	\aboverulesep=.4ex
	\abovetopsep=0pt
	\cmidrulesep=\doublerulesep
	\cmidrulekern=.5em
	\defaultaddspace=.5em
}

\graphicspath{{./figures/}}
\usepackage{xifthen}
\usepackage{xcolor}

\newcommand{\gettitle}{Critical net-proton number fluctuations with hydrodynamics}

\newcommand{\getDalianAffiliation}{\affiliation{School of Physics, Dalian University of Technology, Dalian, 116024, P.R. China}}

\newcommand{\getGiessenAffiliation}{\affiliation{Institut f\"ur Theoretische Physik, Justus-Liebig-Universit\"at Gie\ss en, 35392 Gie\ss en, Germany}}

\newcommand{\getUCBAffiliation}{\affiliation{Department of Physics, University of California, Berkeley, 94270, CA, USA}}

\newcommand{\getLBNLAffiliation}{\affiliation{Nuclear Science Division, Lawrence Berkeley National Laboratory, Berkeley, 94270, CA, USA}}

\newcommand{\getCCNUAffiliation}{\affiliation{Key Laboratory of Quark \& Lepton Physics (MOE) and Institute of Particle Physics, Central China Normal University, Wuhan 430079, China}}

\hypersetup{
	colorlinks,
	linkcolor={blue!75!black},
	citecolor={blue!75!black},
	urlcolor={blue!75!black}, 
	pdftitle={\gettitle},
	pdfauthor={},
	pdfkeywords={}
	{} 
	bookmarksopen=true,
	bookmarksopenlevel=2,
	bookmarksnumbered=true
}

\begin{document}
\title{\gettitle}

\author{Rui-zhe Zhao}
\email{zhaoruizhe@mail.dlut.edu.cn}
\getDalianAffiliation

\author{Shi Yin}
\email{shiyin.dalian@gmail.com}
\getGiessenAffiliation

\author{Shanjin Wu}
\email{shanjinwu@dlut.edu.cn}
\getDalianAffiliation

\author{Lipei Du}
\email{ldu2@lbl.gov}
\getUCBAffiliation
\getLBNLAffiliation

\author{Xiaofeng Luo}
\email{xfluo@ccnu.edu.cn}
\getCCNUAffiliation

\author{Wei-jie Fu}
\email{wjfu@dlut.edu.cn}
\getDalianAffiliation

\begin{abstract}

We compute the net-proton number fluctuations and their ratios $C_2/C_1$, $C_3/C_2$ and $C_4/C_2$ on the hydrodynamic freeze-out hypersurface of particlization at nine collision energies, $\sqrt{s_{\mathrm{NN}}}=7.7-200$ GeV, based on the fluctuations obtained from the functional renormalization group (fRG) approach, where both the regular and the critical fluctuations arising from the critical end point (CEP) are included. The transverse momentum and rapidity acceptance windows as same as the experimental measurements, the isospin randomization for the proton number fluctuations, and the global baryon conservation effect are implemented in the calculations. The results are also compared with the baseline results without critical fluctuations. It is found that for the low-order cumulants, e.g., $C_2/C_1$ the difference between the critical and non-critical results is small, while the difference increases with the increasing order of cumulants in the region of low collision energy. A non-monotonic dependence on the collision energy is observed in $C_4/C_2$ with critical fluctuations, which is absent in the results without critical fluctuations.

\end{abstract}

\maketitle

%%%%%%%%%%%%%%%%%%%%%%%%%%%%%%%%%%%%%%%%%%%%%%%%%%%%%%%%%%%
%%%%%%%%%%%%%%%%%%%%%%%%%%%%%%%%%%%%%%%%%%%%%%%%%%%%%%%%%%%
\emph{Introduction.--} 
QCD phase diagram constitutes a fundamental subject of physics, situated at the cutting-edge frontier of interdisciplinary research, e.g., the quark-gluon plasma (QGP) produced in relativistic heavy-ion collisions within nuclear physics, the properties of compact astrophysical objects including neutron stars, the evolution of the early universe in cosmology, etc. \cite{Stephanov:2007fk, Luo:2017faz, Bzdak:2019pkr, Fu:2022gou, Fukushima:2025ujk, Fischer:2026uni}. The most fascinating feature of the QCD phase diagram is the QCD critical point, also referred to as the critical end point (CEP). It terminates the line of first-order phase transitions at high baryon densities and links to the smooth chiral crossover in the region of low baryon densities. The continuous crossover has long been confirmed in lattice QCD simulations, cf., e.g., \cite{Aoki:2006we}.

The existence and location of the CEP remain an outstanding open question. Direct QCD simulations on lattice at finite baryon chemical potential are severely hindered by the notorious sign problem. Even so, substantial advances in both theoretical and experimental investigations of the CEP have been achieved over recent years. Different theoretical approaches, e.g., the functional QCD, including the functional renormalization group (fRG) \cite{Fu:2019hdw, Braun:2023qak, Fu:2024rto, Pawlowski:2025jpg, Fu:2026qnl, Wang:2026xwa} and Dyson-Schwinger equation (DSE) \cite{Gao:2020fbl, Gunkel:2021oya, Lu:2025cls, Lu:2026ezr}, which are important nonperturbative methods to first-principles QCD and are complementary to lattice QCD, the lattice QCD extrapolation based on the Yang-Lee edge singularities \cite{Basar:2023nkp, Clarke:2024ugt, Adam:2025phc} or contours of constant entropy density \cite{Shah:2024img, Borsanyi:2025dyp}, Bayesian holography \cite{Cai:2022omk, Hippert:2023bel, Zhu:2025gxo}, etc., have arrived at a convergent result that if there is a CEP, its location is favored in a region of relatively large baryon chemical potential, with the ratio of the baryon chemical potential of CEP over its temperature $\mu_B/T \gtrsim 4 \sim 5$.

It has been proposed long time ago that the fluctuation observables can be used to search for the CEP in heavy-ion collision experiments due to the essence of critical phenomenon of the CEP \cite{Stephanov:2008qz, Stephanov:2011pb}. Proton number fluctuations, as the proxy of the baryon number fluctuations, have been measured at the Relativistic Heavy Ion Collider (RHIC) by the STAR collaboration \cite{STAR:2020tga, STAR:2021fge, STAR:2022vlo, STAR:2022etb, STAR:2025zdq}, and a deviation from non-critical-point model calculations in the kurtosis of net-proton number fluctuations  is found \cite{STAR:2025zdq}.

In order to pin down the location of CEP from experimental measurements if it exists, a reliable prediction of experimental fluctuation signals based on theoretical calculations with a CEP is indispensable, which, however, is quite challenging for the moment. A simplified alternative is to compute the critical fluctuations on a parametrized chemical freeze-out curve \cite{Fu:2016tey, Fu:2021oaw, Fu:2023lcm, Karthein:2025hvl, Lu:2026ezr}, where one collision energy corresponds to a single baryon chemical potential and temperature. This is obviously an over-simplified picture, which is far away from the realistic heavy-ion collisions. In this letter, we would like to combine the fRG calculations with the hydrodynamic simulations, which provides a promising approach to describe the fluctuation observables in experiments. This approach can also be improved progressively in the future, with the advance of CEP-related studies or hydrodynamic simulations.

In this letter we first discuss the method and setup: How the baryon number fluctuations with critical fluctuations, calculated in fRG, are encoded in different fluid elements of the particlization hypersurface at the freeze-out obtained from hydrodynamic simulations. Then numerical results are presented. A non-monotonic dependence of the kurtosis of net-proton fluctuations on the collision energy is found in the calculation with fRG and hydrodynamics, in contrast to the counterpart with non-critical fluctuations.

%
%%%%%%%%%%%%%%%%%%%%%%%%%%%%%
\begin{figure}[t]
\includegraphics[width=0.48\textwidth]{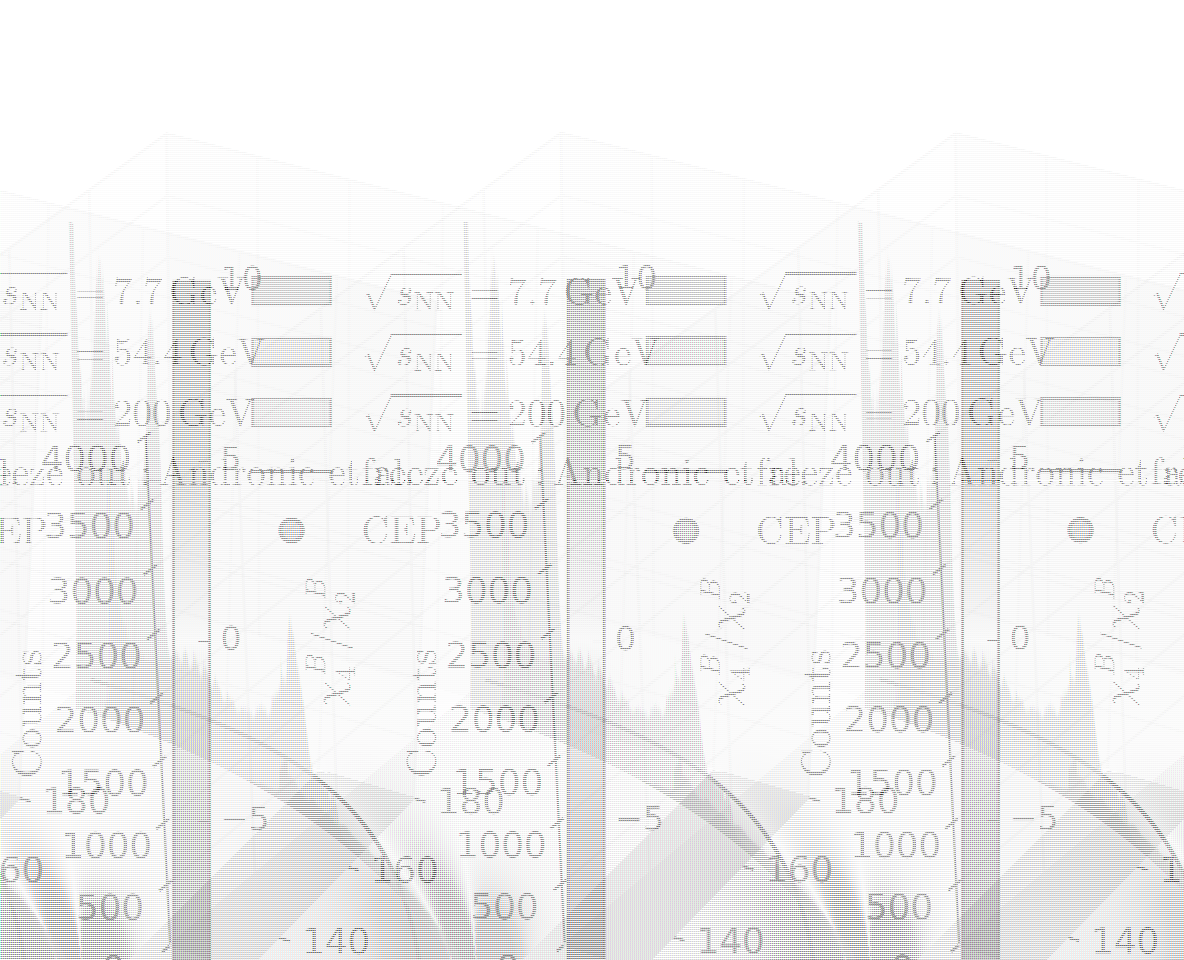}
\caption{Counts of fluid elements distributed on the QCD phase diagram at the freeze-out hypersurface of particlization for three representative collision energies $\sqrt{s_{\mathrm{NN}}}=7.7$ GeV (green), 54.4 GeV (orange), 200 GeV (purple). On the $T$ ($y$ axis)-$\mu_B$ ($x$ axis) plane of phase diagram, the grand canonical kurtosis of net-baryon fluctuations $\chi^B_4/\chi^B_2$ calculated in fRG \cite{Fu:2023lcm} is depicted as a colored heatmap. The red circle indicates the location of CEP. The black solid line in the phase diagram denotes the parametrized chemical freeze-out curve obtained by Andronic et al. \cite{Andronic:2017pug}. The gray area depicts the region where $\chi^B_4/\chi^B_2$ is not computed.}
\label{fig:phasediagram-hydro-3d}
\end{figure}
%%%%%%%%%%%%%%%%%%%%%%%%%%%%%
%

%%%%%%%%%%%%%%%%%%%%%%%%%%%%%%%%%%%%%%%%%%%%%%%%%%%%%%%%%%%
%%%%%%%%%%%%%%%%%%%%%%%%%%%%%%%%%%%%%%%%%%%%%%%%%%%%%%%%%%%
\emph{Critical fluctuations encoded on the freeze-out hypersurface of particlization from hydrodynamics.--} 
In this letter, we evolve the hydrodynamic simulations of heavy-ion collisions until the freeze-out hypersurface of particlization, where the hypersurface is comprised of many elements $\delta\sigma_\mu(x_i)$, characterized by different temperature $T(x_i)$, baryon chemical potential $\mu_B(x_i)$, and the 4-velocity $u^\mu(x_i)$, etc. Here $x_i$ distinguishes different fluid elements. The net-baryon number fluctuations including critical fluctuations related to the CEP calculated from the fRG \cite{Fu:2023lcm} are employed to determine the cumulants of baryon number fluctuations for each fluid element. This approach combines the fRG with the hydrodynamics, which allows us to study the critical fluctuations of baryon or proton numbers in a realistic environment close to heavy-ion collisions, where the transverse momentum and rapidity acceptance cuts in experiments, isospin randomization for the proton fluctuations, the global baryon  conservation effect, etc., are readily implemented.

The spacetime evolution of Au+Au collisions is simulated using a state-of-the-art (3+1)-dimensional multistage hydrodynamic framework with rapidity-dependent three-dimensional initial conditions, MUSIC viscous hydrodynamics~\cite{Schenke:2010nt, Schenke:2011bn, Paquet:2015lta}, iS3D particlization~\cite{McNelis:2019auj}, and the UrQMD hadronic afterburner~\cite{Bass1998,Bleicher1999}. The hydrodynamic stage evolves the energy-momentum tensor and net baryon current, including shear-stress and baryon-diffusion effects~\cite{Denicol2018,Du:2019obx,Shen:2020jwv}, until particlization on a constant-energy-density hypersurface at $\varepsilon_{\rm fo}=0.26 \, {\rm GeV} / {\rm fm}^3$.

The model parameters have been systematically calibrated to hadronic observables at nine beam energies, $\sqrt{s_{\mathrm{NN}}}=7.7-200$ GeV, including identified hadron yields, transverse momentum spectra, mean transverse momenta, and rapidity distributions~\cite{Du:2022yok,Du:2023gnv,Du:2023efk}. The same dynamical framework has also been used to calculate thermal photon and dilepton spectra, providing a consistent description of available electromagnetic measurements across this energy range~\cite{Churchill:2023vpt,Churchill:2023zkk,Du:2024pbd,Du:2025dot}. It therefore provides a well-constrained dynamical background for the present study of critical fluctuations.

In \Fig{fig:phasediagram-hydro-3d} we show the counts of fluid elements at the freeze-out hypersurface, distributed on the QCD phase diagram in terms of $T$ and $\mu_B$. Three representative collision energies $\sqrt{s_{\mathrm{NN}}}=7.7$, 54.4, 200 GeV are chosen. On the phase diagram we also plot the grand canonical kurtosis of net-baryon fluctuations $\chi^B_4/\chi^B_2$ calculated in fRG \cite{Fu:2023lcm}, where $\chi^B_n$ denotes the generalized susceptibilities of baryon number, i.e.
\begin{align}
    \chi_{n}^{B}=\frac{\partial^{n}}{\partial(\mu_{B}/T)^{n}}\frac{p}{T^{4}}\,, \label{eq:chi-def}
\end{align}
with the pressure $p$. The red circle on the phase diagram stands for the location of CEP obtained in fRG \cite{Fu:2023lcm}, which reads 
\begin{align}
    (T_{_\text{CEP}},\mu_{B_{\text{CEP}}})=(98,643)\,\text{MeV}\,. \label{eq:CEP}
\end{align}
Note that the location of CEP in \Eq{eq:CEP} is consistent with recent results of QCD from fRG \cite{Fu:2019hdw, Fu:2026qnl, Wang:2026xwa} and DSE \cite{Gao:2020fbl, Gunkel:2021oya}, as well as lattice QCD extrapolation \cite{Basar:2023nkp, Clarke:2024ugt, Shah:2024img, Borsanyi:2025dyp}.

Different from the parametrized chemical freeze-out curve as shown by the black solid line in \Fig{fig:phasediagram-hydro-3d}, where one collision energy corresponds to one value of freeze-out $\mu_B$ or $T$, the particlization hypersurface of hydrodynamics for each collision energy is composed of fluid elements of different $\mu_B$ and $T$. One can see from \Fig{fig:phasediagram-hydro-3d} that the elements of $\sqrt{s_{\mathrm{NN}}}=7.7$, 54.4, 200 GeV are  distributed in a narrow band of $15\,\text{MeV}\lesssim\mu_B \lesssim 130 \,\text{MeV}$, $45\,\text{MeV}\lesssim\mu_B \lesssim 430 \,\text{MeV}$, $220\,\text{MeV}\lesssim\mu_B \lesssim 600 \,\text{MeV}$, respectively, see \cite{Du:2023gnv} for more details. The number of elements decreases with the decreasing collision energy.  

For each fluid element on the freeze-out surface obtained from hydrodynamic simulations, we assume the fluctuations all reach thermal equilibrium and independent from other elements. The $n$-th order fluctuations of baryon or antibaryon number of fluid cell $i$ in the grand-canonical ensemble read
\begin{align}
    \delta C_n^{B^\pm,\mathrm{gce}}(x_i)=\delta V_i \big(T(x_i)\big)^3\chi_n^{B^\pm}(x_i)\,, \label{eq:CBpm}
\end{align}
with the effective volume $\delta V_i =\delta \sigma_\mu(x_i) u^\mu(x_i)$ for the hypersurface element $x_i$. Here $\chi_n^{B^\pm}$ represents the $n$-th order susceptibilities of baryon or antibaryon number fluctuations, respectively. Note that theoretical calculations of critical fluctuations from, e.g., lattice QCD or fRG, can only provide the net-baryon number fluctuations, here denoted by $\chi_n^{B}$, rather than fluctuations for the baryon or antibaryon numbers separately. In this letter we adopt the relation as follows
\begin{align}
    \chi_n^{B^\pm}=\frac{\chi_n^{B^\pm,\text{HRG}}}{\chi_n^{B,\text{HRG}}}\chi_n^{B}\,,\label{eq:chiBpm-chiB}
\end{align}
for each fluid element, such that the critical (anti)baryon number fluctuations can be inferred indirectly from the critical net-baryon fluctuations obtained from fRG \cite{Fu:2023lcm}, in each cell with temperature $T(x_i)$ and baryon chemical potential $\mu_B(x_i)$. In \Eq{eq:chiBpm-chiB} $\chi_n^{B^\pm,\text{HRG}}$ and $\chi_n^{B,\text{HRG}}$ stand for the non-critical (anti)baryon and net-baryon number fluctuations in the hadron resonance gas (HRG), respectively,  which both can be readily computed at the same $T(x_i)$ and $\mu_B(x_i)$.

To take into account the effects of finite momentum acceptance of the detector, we employ the method of binomial acceptance corrections~\cite{Savchuk:2019xfg, Vovchenko:2021kxx}. The probability to detect an (anti)baryon within a momentum acceptance ${\Delta p_{\mathrm{acc}}}$ can be obtained from the Cooper-Frye formula for the particlization at the freeze-out hypersurface \cite{Cooper:1974mv}, i.e.,  
\begin{align}
    &\quad p_\mathrm{acc}(x_i; \Delta p_\mathrm{acc})\nonumber \\[2ex]
    &=\frac{\int_{p\in\Delta p_\mathrm{acc}}\frac{d^3p}{\omega_p}\delta\sigma_\mu(x_i)p^\mu f\big[u^\nu(x_i)p_\nu; T(x_i),\mu_j(x_i)\big]}{\int\frac{d^3p}{\omega_p}\delta\sigma_\mu(x_i)p^\mu f\big[u^\nu(x_i)p_\nu; T(x_i),\mu_j(x_i)\big]}\,,\label{eq:pacc}
\end{align}
with the Boltzmann distribution 
\begin{align}
    f[u^\mu p_\mu; T(x),\mu_j(x)]=\frac{d_j}{(2\pi)^3}\exp\left[\frac{\mu_j(x)-u^\mu(x)p_\mu}{T(x)}\right]\,,
\end{align}
where $d_j$ and $\mu_j $ denote the degeneracy and chemical potential for the (anti)baryon species $j$, respectively. In the actual calculations of $p_\mathrm{acc}$ in \Eq{eq:pacc}, all (anti)baryons are assumed to have the same nucleon mass $m_N=0.938$ GeV. Then, (anti)baryon number cumulants of different orders observed in the acceptance window read
\begin{align}
    &\quad \delta C_n^{B^\pm,\mathrm{gce}}(x_i; \Delta p_\mathrm{acc})\nonumber \\[2ex]
    &=\sum_{l=1}^n\delta C_l^{B^\pm,\mathrm{gce}}(x_i)B_{n,l}\left(\phi_t',\cdots,\phi_t^{(n-l+1)}\right)\,,\label{eq:CBpm-pacc}
\end{align}
with the partial Bell polynomials $B_{n,l}$ and $\phi(t)=\ln(1-p_\mathrm{acc}+e^t p_\mathrm{acc})$, where the generating variable $t$ as well as $t_p$ in the following is chosen to be $t=t_p=0$ finally in the calculations. The momentum cut $\Delta p_{\mathrm{acc}}$ in this letter is chosen to being consistent with the colliding mode of STAR detector: $0.4<p_T<2.0$ GeV for the transverse momentum and $|y|<0.5$ for the rapidity. The dependence of results on the varying acceptance of transverse momentum and rapidity is studied in detail in the supplement.

%
%%%%%%%%%%%%%%%%%%%%%%%%%%%%%
\begin{figure*}[t]
\includegraphics[width=0.33\textwidth]{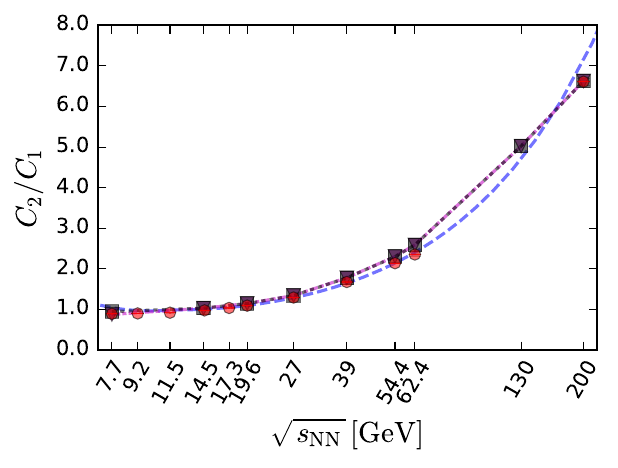}%\hspace{0.1cm}
\includegraphics[width=0.33\textwidth]{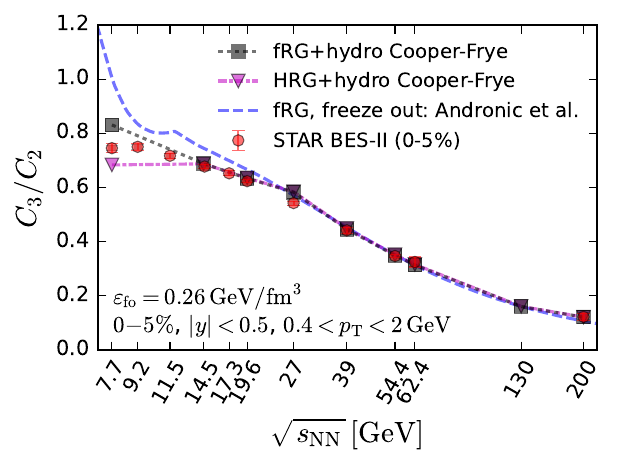}%\hspace{0.1cm}
\includegraphics[width=0.33\textwidth]{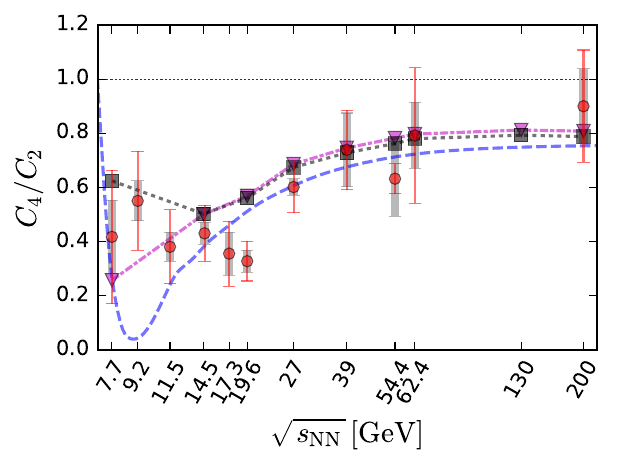}
\caption{Collision energy dependence of net-proton cumulant ratios $C_2/C_1$ (left), $C_3/C_2$ (middle) and $C_4/C_2$ (right) in $0-5\%$ Au-Au collisions at $\sqrt{s_{\mathrm{NN}}}=7.7-200$ GeV. The black squares represent the results obtained with critical fluctuations from fRG combined with the hydrodynamic freeze-out Cooper-Frye hypersurface, incorporating acceptance corrections, isospin randomization, and the effects of global baryon conservation. The magenta inverted triangles show the non-critical baseline results obtained within the same framework but with the susceptibilities taken from the HRG model. The blue dashed line stands for the fRG results \cite{Fu:2023lcm}, computed along a simple parametrized chemical freeze-out curve obtained by Andronic et al. \cite{Andronic:2017pug}. The red circles denote the experimental data of the phase II of Beam Energy Scan program (BES-II) at RHIC by the STAR collaboration \cite{STAR:2025zdq}.}
\label{fig:cumulants}
\end{figure*}
%%%%%%%%%%%%%%%%%%%%%%%%%%%%%
%

In order to calculate the (anti)proton number fluctuations, we adopt the argument of isospin randomization in \cite{Kitazawa:2011wh, Kitazawa:2012at}, such that the (anti)proton number fluctuations can be obtained via the binomial filtering of the (anti)baryon number fluctuations. The joint cumulants of baryon-proton or antibaryon-antiproton distribution read 
\begin{align}
    &\delta C_{n,m}^{B^\pm,p^\pm,\mathrm{gce}}(x_i;\Delta p_\mathrm{acc})=\delta_{m,0}\delta C_n^{B^\pm,\mathrm{gce}}(x_i;\Delta p_\mathrm{acc})\nonumber \\[2ex]
    &\quad+\sum_{l=1}^m\delta C_{n+l}^{B^\pm,\mathrm{gce}}(x_i;\Delta p_\mathrm{acc})B_{m,l}\left(\gamma_{t_p}',\cdots,\gamma_{t_p}^{(m-l+1)}\right)\,,\label{eq:CBp}
\end{align}
with $\gamma(t_p,q)= \ln[1-(1-e^{t_p})q]$, where $q$ represents the probability that a randomly chosen (anti)baryon is a (anti)proton, and can be evaluated by the ratio between the mean values of (anti)proton and (anti)baryon number, i.e., $q(x_i)= \langle N_{p^\pm}(x_i)\rangle/\langle N_{B^\pm}(x_i)\rangle$ for fluid element $x_i$. It is found that $q(x_i)\approx1/2$ in most cases in the HRG model with weak decay contributions included \cite{Vovchenko:2021kxx}, and thus $q(x_i)=1/2$ is assumed throughout this letter. In \Eq{eq:CBp}, the subscripts $n$ and $m$ stand for the orders of cumulants for the (anti)baryon and (anti)proton, respectively. By summing up \Eq{eq:CBp} over all the fluid elements, one arrives at the total joint cumulants in the acceptance, i.e.,
\begin{align}
    C_{n,m}^{B^\pm,p^\pm,\mathrm{gce}}(\Delta p_\mathrm{acc})=\sum_{i\in\sigma}\delta C_{n,m}^{B^\pm,p^\pm,\mathrm{gce}}(x_i;\Delta p_\mathrm{acc})\,. \label{eq:CBp-sum}
\end{align}
Furthermore, the joint cumulants of net-baryon and net-proton number distributions can be obtained as
\begin{align}
    C_{n,m}^{B,p,\mathrm{gce}}(\Delta p_\mathrm{acc})&=C_{n,m}^{B^+,p^+,\mathrm{gce}}(\Delta p_\mathrm{acc})\nonumber \\[2ex]
    &\quad +(-1)^{n+m} C_{n,m}^{B^-,p^-,\mathrm{gce}}(\Delta p_\mathrm{acc})\,. \label{eq:CBp-net}
\end{align}
In particular, $C_{0,m}^{B,p,\mathrm{gce}}(\Delta p_\mathrm{acc})$ with $n=0$ corresponds to the $m$-th order cumulants of the net-proton number distributions within the momentum acceptance. Note that in \Eq{eq:CBp-net} the correlations between the baryons and antibaryons are neglected. In fact, \Eq{eq:chiBpm-chiB} and \Eq{eq:CBp-net} should be put together and regarded as an assumption on the relation between the net-baryon and the (anti)baryon for the critical fluctuations, since only the net-baryon critical fluctuations are genuinely computable from theory. This is attributed to an obvious reason that the net-baryon is a conserved charge, while the (anti)baryon is not. The reasonability of the assumption in \Eq{eq:chiBpm-chiB} and \Eq{eq:CBp-net} is partly reflected by the observation that if all the acceptance cut is removed, the critical fluctuations are exact even after the implementation of \Eq{eq:chiBpm-chiB} and \Eq{eq:CBp-net} is done.

In order to take into account the effects of global baryon conservation on the cumulants of the net-baryon or net-proton in the acceptance window, i.e., the canonical ensemble (ce) corrections, we adopt the subensemble acceptance method (SAM) \cite{Vovchenko:2020tsr}, further developed in \cite{Vovchenko:2021yen}. The cumulants with the canonical ensemble corrections read
\begin{align}
    C_{n,m}^{B,p,\mathrm{ce}}(\Delta p_\mathrm{acc})=\mathcal{S}\left[C_{n,m}^{B,p,\mathrm{gce}}(\Delta p_\mathrm{acc}),C_{n,m}^{B,p,\mathrm{gce}}(\overline{\Delta p_\mathrm{acc}})\right]\,, \label{eq:CBp-SAM}
\end{align}
where $\overline{\Delta p_\mathrm{acc}}$ denotes the momentum outside the acceptance. The cumulants of baryons or protons outside the acceptance can be calculated similarly as those in the acceptance just with the probability in \Eq{eq:pacc} replaced with $p_\mathrm{acc}(x_i; \overline{\Delta p_\mathrm{acc}})=1-p_\mathrm{acc}(x_i; \Delta p_\mathrm{acc})$. The mapping $\mathcal{S}$ in \Eq{eq:CBp-SAM} is obtained in \cite{Vovchenko:2021yen} and also collected in the supplement.

%%%%%%%%%%%%%%%%%%%%%%%%%%%%%%%%%%%%%%%%%%%%%%%%%%%%%%%%%%%
%%%%%%%%%%%%%%%%%%%%%%%%%%%%%%%%%%%%%%%%%%%%%%%%%%%%%%%%%%%
\emph{Results and discussion.--}
In \Fig{fig:cumulants} we show the variance $\sigma^2/C_1=C_2/C_1$, skewness $S\sigma=C_3/C_2$ and kurtosis $\kappa\sigma^2 = C_4/C_2$ of the net-proton distributions at collision energy $\sqrt{s_{\mathrm{NN}}}=7.7-200$ GeV, obtained with the critical fluctuations from fRG combined with the hydrodynamic freeze-out Cooper-Frye hypersurface, where the momentum acceptance, isospin randomization for the proton number fluctuations from the baryon number fluctuations, and the global baryon conservation are taken into account.  Moreover, two other results are also presented for comparison: One is the non-critical baseline result obtained with the HRG fluctuations computed on the hydrodynamic hypersurface. The other is the fRG result with critical fluctuations but without the hydrodynamic hypersurface, which was obtained in \cite{Fu:2023lcm}. In \cite{Fu:2023lcm} the fRG fluctuations were calculated along a parametrized chemical freeze-out curve obtained in \cite{Andronic:2017pug}, where one value of collision energy corresponds to a single value of baryon chemical potential or temperature at the chemical freeze-out. Furthermore, in \cite{Fu:2023lcm} no momentum cut was applied and only the net-baryon number fluctuations were calculated, whose ratios were assumed to be the same as the net-proton cumulant ratios. The global baryon conservation effect was also included via the SAM in \cite{Fu:2023lcm}.

Note that in \Fig{fig:cumulants} the cumulant ratios of fRG or HRG combined with the hydrodynamic hypersurface are obtained in $0-5\%$ central Au-Au collisions, with the rapidity acceptance $|y|<0.5$ and the transverse momentum $0.4< p_T < 2$ GeV, being consistent with the colliding mode of STAR experiment \cite{STAR:2025zdq}. The constant switching energy density at the freeze-out hypersurface of particlization is chosen to be $\varepsilon_{\rm fo}=0.26 \, {\rm GeV} / {\rm fm}^3$ for all collision energies, determined by fitting bulk observables in hydrodynamic simulations, which is in agreement with the choice in \cite{Vovchenko:2021kxx, Du:2023gnv}. We do not present results at $\sqrt{s_\mathrm{NN}}=9.2$ and $11.5$~GeV, where the available hadronic measurements do not support a comparably controlled calibration of the hydrodynamic background.

Comparing the result of fRG with hydrodynamics with that of HRG with hydrodynamics in \Fig{fig:cumulants}, i.e., the black squares and the magenta triangles, one immediately observes that the difference between them is negligible in the whole collision energy range investigated here for the low order cumulants $C_2/C_1$, or in the regime of intermediate and high collision energies, say $\sqrt{s_{\mathrm{NN}}}\gtrsim 14.5$ GeV, for the high order cumulants $C_3/C_2$ and $C_4/C_2$, indicating the impact of critical fluctuations there is not important, while the cumulant ratios are mainly determined by the non-critical fluctuations, the acceptance window and the related global baryon conservation. For instance, the deviation of $C_4/C_2$ from the unity in the region of high collision energy is mainly attributed to the effect of global baryon conservation. This is reasonable, since low order cumulants carry less critical fluctuations and high collision energies are related to the region of small baryon chemical potential in the QCD phase diagram, where the critical fluctuations are small. However, the situation is quite different in the regime of low collision energy, which probes the region of high baryon chemical potential in the QCD phase diagram, as shown in \Fig{fig:phasediagram-hydro-3d}. It is found at $\sqrt{s_{\mathrm{NN}}}= 7.7$ GeV there is some difference in $C_3/C_2$ between the fRG with hydrodynamics and the HRG with hydrodynamics. The difference in $C_4/C_2$ becomes more pronounced, and a non-monotonic behavior is observed for the fRG with hydrodynamics while this is not observed in the results of HRG with hydrodynamics. Moreover, in the supplement it is found that the non-monotonic behavior in $C_4/C_2$, obtained from fRG combined with hydrodynamics, persists with varying transverse momentum and rapidity acceptance cuts.

In the same way, it is interesting and valuable to compare the fRG results with and without the hydrodynamics in \Fig{fig:cumulants}. Although critical fluctuations are included in both calculations, it seems that the blue dashed line overestimates the non-monotonicity in $C_4/C_2$ in comparison to the experimental data, which indicates that the simplified picture of evaluating fluctuations on a simple chemical freeze-out curve is still far away from the actual heavy-ion collisions. On the contrary, the fRG combined with hydrodynamics yields more comparable results in both $C_3/C_2$ and $C_4/C_2$, in comparison to the experimental data. In $C_2/C_1$ the three theoretical curves are consistent with experimental data.

%%%%%%%%%%%%%%%%%%%%%%%%%%%%%%%%%%%%%%%%%%%%%%%%%%%%%%%%%%%%%%%%%%%%%%%%%%%%%%%%%%%%%%%%%%%%%%%%%%%%%%%%%%%%%%%%%%%%%%%
\emph{Conclusions.--}
In this letter, we evolve the hydrodynamic simulations of heavy-ion collisions at nine collision energies, $\sqrt{s_{\mathrm{NN}}}=7.7-200$ GeV, to the freeze-out hypersurface of particlization. Then the net-baryon number cumulants for the different fluid elements on the hypersurface are computed via the fRG approach, where both the regular and the critical fluctuations arising from the CEP are included. Then, the transverse momentum and rapidity acceptance windows are applied as same as the experimental measurements. The net-proton number fluctuations are computed from the baryon number fluctuations with the method of isospin randomization. The global baryon conservation effect is taken into account through the subensemble acceptance method.

We compute the net-proton number fluctuations of different orders within this framework. The collision energy dependence of net-proton cumulant ratios $C_2/C_1$, $C_3/C_2$ and $C_4/C_2$ is investigated. The results are also compared with the baseline results without critical fluctuations. It is found that for the low-order cumulants, e.g., $C_2/C_1$ the difference between the critical and non-critical results is small, while the difference increases with the increasing order of cumulants in the region of low collision energy. A non-monotonic dependence on the collision energy is observed in $C_4/C_2$ with critical fluctuations, while this is not observed in the case without critical fluctuations. It is very interesting to extend our calculations to the fixed-target energy range with $\sqrt{s_{\mathrm{NN}}}<7.7$ GeV in the future.

%%%%%%%%%%%%%%%%%%%%%%%%%%%%
\emph{Acknowledgements.--}
We thank Fei Gao, Jianing Li, Yu-xin Liu and Jan M. Pawlowski for discussions and comments. We also would like to thank the members of the fQCD collaboration \cite{fQCD} for collaborations on related projects.
W.J. Fu and X.F. Luo are supported by the National Natural Science Foundation of China under Contract No.\ 12447102. 
X.F. Luo is also supported by the National Natural Science Foundation of China under Contract No.\ 12525509, and the National Key Research and Development Program of China under Contract No.\ 2022YFA1604900, and the Fundamental Research Funds for the Central Universities (XJ2026000302).
S.J. Wu is supported by the Fundamental Research Funds for the Central Universities at Dalian University of Technology and the National Natural Science Foundation of China under Contract No.\ 12305143. 
S. Yin is supported by the Alexander von Humboldt foundation.

\bibliography{ref-lib}% Produces the bibliography via BibTeX.

%%%%%%%%%%%%%%%%%%%%%%%%%%%%%%%%%%%%%%%%%%%%%
%\newpage 

\clearpage

%\appendix 
\renewcommand{\thesubsection}{{S.\arabic{subsection}}}
\setcounter{section}{0}
\titleformat*{\section}{\centering \Large \bfseries}

\onecolumngrid

%\begin{widetext}
%	\newcounter{totalequations}
\section*{Supplemental Materials}
%	\parttoc % Insert the appendix TOC

The supplemental materials provide additional results for the studies of the acceptance dependence of the net proton fluctuations calculated with the fRG and hydrodynamics in \sec{app:acceptance}, the canonical ensemble corrections from the subensemble acceptance method in \sec{app:SAM}.

%%%%%%%%%%%%%%%%%%%%%%%%%%%%%%%%%%%%%%%%%%%%%%%%%%%%%%%%%%%%%%%%%%%%%%%%%%%%%%%%%%%%%%%%%%%%%%%%%%%%%%%%%%%%%%%%%%%%%%%
\subsection{Acceptance dependence}
\label{app:acceptance}

%
%%%%%%%%%%%%%%%%%%%%%%%%%%%%%
\begin{figure*}[th]
\includegraphics[width=0.33\textwidth]{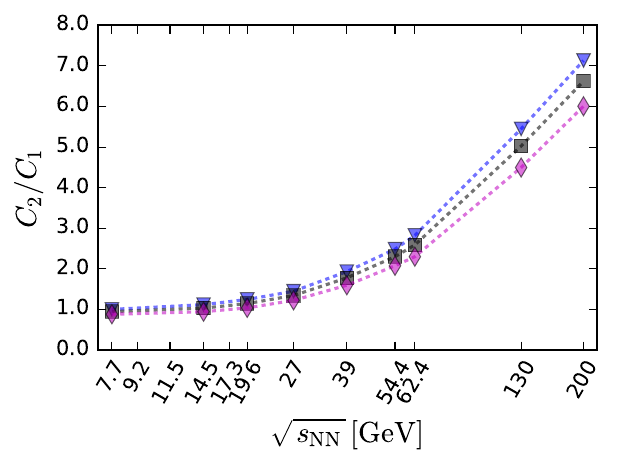}%\hspace{0.1cm}
\includegraphics[width=0.33\textwidth]{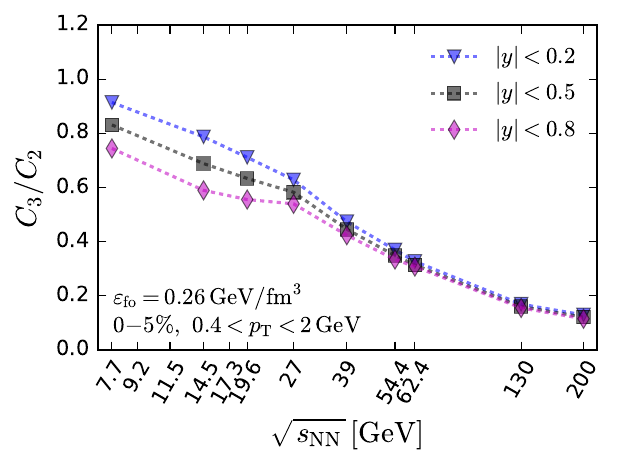}%\hspace{0.1cm}
\includegraphics[width=0.33\textwidth]{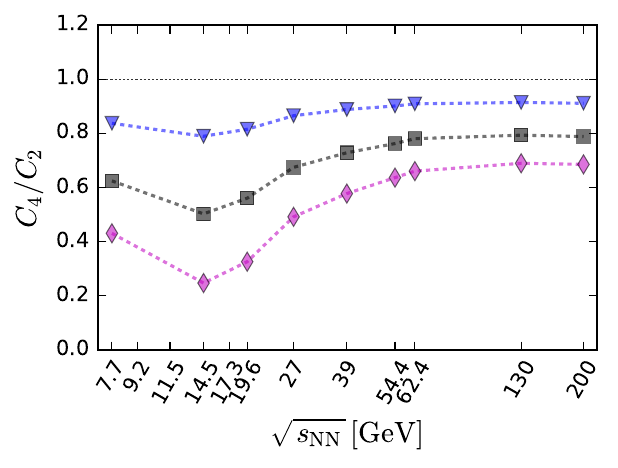}
\caption{Collision energy dependence of net-proton cumulant ratios $C_2/C_1$ (left), $C_3/C_2$ (middle) and $C_4/C_2$ (right) in $0-5\%$ Au-Au collisions at $\sqrt{s_{\mathrm{NN}}}=7.7-200$ GeV, shown for three rapidity acceptance windows: $|y|<0.2$ (blue), $|y|<0.5$ (black), and $|y|<0.8$ (magenta). The transverse momentum acceptance is fixed with $0.4<p_T<2.0$ GeV. All results are obtained using critical fluctuations from fRG combined with the hydrodynamic freeze-out Cooper-Frye hypersurface, incorporating acceptance corrections, isospin randomization, and the effects of global baryon conservation.}
\label{fig:cumulants-y}
\end{figure*}
%%%%%%%%%%%%%%%%%%%%%%%%%%%%%
%

%
%%%%%%%%%%%%%%%%%%%%%%%%%%%%%
\begin{figure*}[th]
\includegraphics[width=0.33\textwidth]{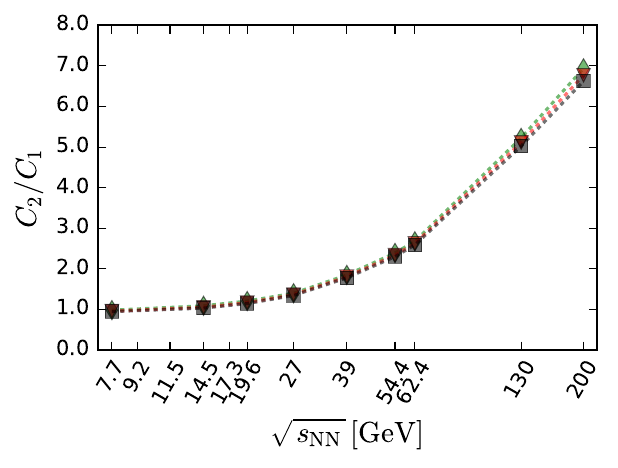}%\hspace{0.1cm}
\includegraphics[width=0.33\textwidth]{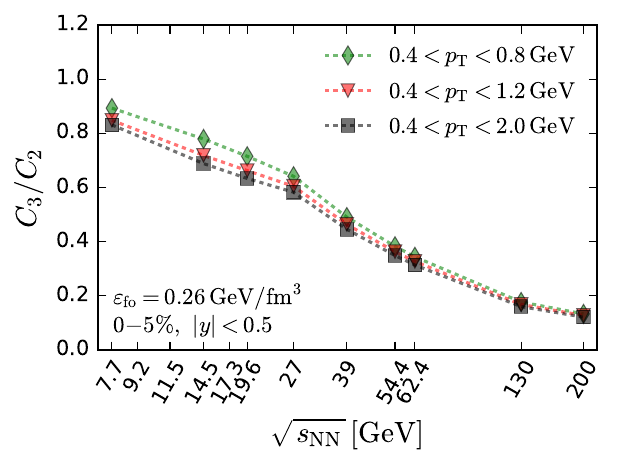}%\hspace{0.1cm}
\includegraphics[width=0.33\textwidth]{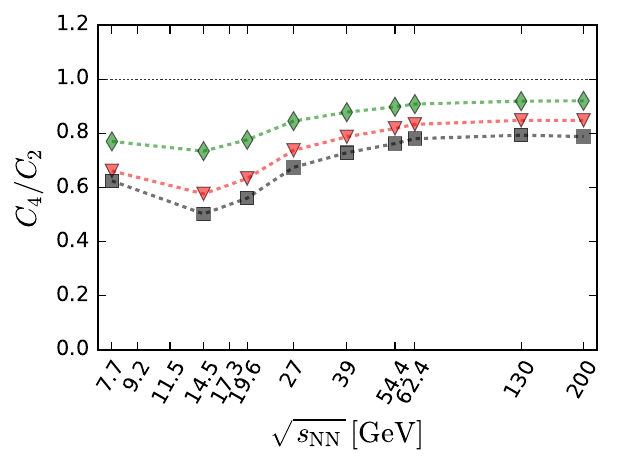}
\caption{Collision energy dependence of net-proton cumulant ratios $C_2/C_1$ (left), $C_3/C_2$ (middle) and $C_4/C_2$ (right) in $0-5\%$ Au-Au collisions at $\sqrt{s_{\mathrm{NN}}}=7.7-200$ GeV, shown for three transverse momentum acceptance windows: $0.4<p_T<0.8$ GeV (green), $0.4<p_T<1.2$ GeV (red), and $0.4<p_T<2.0$ GeV (black). The rapidity acceptance is fixed with $|y|<0.5$. All results are obtained using critical fluctuations from fRG combined with the hydrodynamic freeze-out Cooper-Frye hypersurface, incorporating acceptance corrections, isospin randomization, and the effects of global baryon conservation.}
\label{fig:cumulants-pT}
\end{figure*}
%%%%%%%%%%%%%%%%%%%%%%%%%%%%%
%

In this section we investigate the acceptance dependence of net-proton cumulant ratios calculated from the fRG and hydrodynamics. The results for the different rapidity and transverse momentum cuts are presented in \Fig{fig:cumulants-y} and \Fig{fig:cumulants-pT}, respectively. It is found that the degree of acceptance dependence increases with the order of cumulants, no matter for the rapidity in \Fig{fig:cumulants-y} or the transverse momentum in \Fig{fig:cumulants-pT}, where the dependence of $C_2/C_1$ on the acceptance is weakest while the $C_4/C_2$ is strongest. One can see that with the increase of the acceptance window the cumulant ratios decreases. This is because with the increase of the acceptance window, the global baryon conservation effect becomes more and more prominent, which suppresses the fluctuations \cite{Vovchenko:2021kxx}, see also \cite{Vovchenko:2020tsr, Fu:2023lcm}. Furthermore, the acceptance dependence is more pronounced in the region of low collision energy in comparison to that of high collision energy. It is interesting to find that the non-monotonic dependence of $C_4/C_2$ on the collision energy is observed in all the acceptance windows investigated here, which becomes more obvious when the acceptance window is enlarged.

%%%%%%%%%%%%%%%%%%%%%%%%%%%%%%%%%%%%%%%%%%%%%%%%%%%%%%%%%%%%%%%%%%%%%%%%%%%%%%%%%%%%%%%%%%%%%%%%%%%%%%%%%%%%%%%%%%%%%%%
\subsection{Canonical ensemble corrections from the subensemble acceptance method (SAM)}
\label{app:SAM}

The relation between the cumulants of net-baryon and net-proton distributions with canonical ensemble corrections and the grand canonical ensemble cumulants is given by the mapping $\mathcal{S}$ \cite{Vovchenko:2021yen}, as follows
\begin{align}
    C_{n,m}^{B,p,\mathrm{ce}}(\Delta p_\mathrm{acc})=\mathcal{S}\left[C_{n,m}^{B,p,\mathrm{gce}}(\Delta p_\mathrm{acc}),C_{n,m}^{B,p,\mathrm{gce}}(\overline{\Delta p_\mathrm{acc}})\right]\,. \label{eq:CBp-SAM-app}
\end{align}
Since we are only interested in the net-proton cumulants, we consider the cumulants of $n=0$ and first four orders $m$ in the left side of \Eq{eq:CBp-SAM-app}, which read
\begin{align}
    C_{0,1;\mathrm{in}}^\mathrm{ce}&=C_{0,1;\mathrm{in}}^\mathrm{gce}\,, \\[2ex]
    C_{0,2;\mathrm{in}}^\mathrm{ce}&=C_{0,2;\mathrm{in}}^\mathrm{gce}-\frac{\left(C_{1,1;\mathrm{in}}^\mathrm{gce}\right)^2}{\mathcal{A}}\,,\\[2ex]
    %%%%%%
    C_{0,3;\mathrm{in}}^\mathrm{ce}&=C_{0,3;\mathrm{in}}^\mathrm{gce}-\frac{3C_{1,1;\mathrm{in}}^\mathrm{gce}C_{1,2;\mathrm{in}}^\mathrm{gce}}{\mathcal{A}}+\frac{3\left(C_{1,1;\mathrm{in}}^\mathrm{gce}\right)^2C_{2,1;\mathrm{in}}^\mathrm{gce}}{\mathcal{A}^2}-\frac{\mathcal{B}\left(C_{1,1;\mathrm{in}}^\mathrm{gce}\right)^3}{\mathcal{A}^3}\,,\\[2ex]
    %%%%%%
    C_{0,4;\mathrm{in}}^\mathrm{ce}&=C_{0,4;\mathrm{in}}^\mathrm{gce}-\frac{4C_{1,1;\mathrm{in}}^\mathrm{gce}C_{1,3;\mathrm{in}}^\mathrm{gce}+3\left(C_{1,2;\mathrm{in}}^\mathrm{gce}\right)^2}{\mathcal{A}}+\frac{6C_{1,1;\mathrm{in}}^\mathrm{gce}\left(C_{1,1;\mathrm{in}}^\mathrm{gce}C_{2,2;\mathrm{in}}^\mathrm{gce}+2C_{1,2;\mathrm{in}}^\mathrm{gce}C_{2,1;\mathrm{in}}^\mathrm{gce}\right)}{\mathcal{A}^2}\nonumber\\[2ex]
    &\quad-\frac{2\left(C_{1,1;\mathrm{in}}^\mathrm{gce}\right)^2\left[2C_{1,1;\mathrm{in}}^\mathrm{gce}C_{3,1;\mathrm{in}}^\mathrm{gce}+3\mathcal{B}C_{1,2;\mathrm{in}}^\mathrm{gce}+6\left(C_{2,1;\mathrm{in}}^\mathrm{gce}\right)^2\right]}{\mathcal{A}^3}+\frac{\left(C_{1,1;\mathrm{in}}^\mathrm{gce}\right)^3\left(\mathcal{C} C_{1,1;\mathrm{in}}^\mathrm{gce}+12\mathcal{B}C_{2,1;\mathrm{in}}^\mathrm{gce}\right)}{\mathcal{A}^4}\nonumber\\[2ex]
    &\quad-\frac{3\mathcal{B}^2\left(C_{1,1;\mathrm{in}}^\mathrm{gce}\right)^4}{\mathcal{A}^5}\,,
\end{align}
with 
\begin{align}
    \mathcal{A}&=C_{2,0;\mathrm{in}}^\mathrm{gce}+C_{2,0;\mathrm{out}}^\mathrm{gce},\nonumber\\
    \mathcal{B}&=C_{3,0;\mathrm{in}}^\mathrm{gce}+C_{3,0;\mathrm{out}}^\mathrm{gce},\nonumber\\
    \mathcal{C}&=C_{4,0;\mathrm{in}}^\mathrm{gce}+C_{4,0;\mathrm{out}}^\mathrm{gce}.
\end{align}
where the variables with the subscripts $_\mathrm{in}$ and $_\mathrm{out}$ denote those related to $\Delta p_\mathrm{acc}$ and $\overline{\Delta p_\mathrm{acc}}$ in \Eq{eq:CBp-SAM-app}, respectively.

\end{document}